\documentclass[12pt]{article}
\def \tl{\tilde{\lambda}}

\def \cn{Collaboration}

\newcommand{\thspace}{\kern.08333em}

\def \beq{\begin{equation}}
\def \eeq{\end{equation}}
\def \beqn{\begin{eqnarray}}
\def \eeqn{\end{eqnarray}}

\def \s{\sqrt{2}}

\def \v#1#2{V_{#1#2}}

\begin{document}
\rightline{TECHNION-PH-00-25}
\rightline{EFI-2000-8}
\rightline{hep-ph/0003119}
\rightline{March 2000}
\bigskip
\bigskip
\centerline{{\bf THE ROLE OF $B_s \to K \pi$ IN DETERMINING THE WEAK PHASE
 $\gamma$}
\footnote{To be submitted to Physics Letters B.}}
\bigskip
\centerline{\it Michael Gronau}
\centerline{\it Technion -- Israel Institute of Technology, 32000 Haifa,
 Israel}
\medskip
\centerline{and}
\medskip
\centerline{\it Jonathan L. Rosner}
\centerline{\it Enrico Fermi Institute and Department of Physics}
\centerline{\it University of Chicago, Chicago, IL 60637}
\bigskip
\centerline{\bf ABSTRACT}
\medskip
\begin{quote}
The decay rates for $B^0 \to K^+ \pi^-$, $B^+ \to K^0 \pi^+$, and the
charge-conjugate processes were found to provide information on the weak phase
$\gamma \equiv {\rm Arg}(V_{ub}^*)$ when the ratio $r$ of weak tree and penguin
amplitudes was taken from data on $B \to \pi \pi$ or semileptonic $B \to \pi$
decays.  We show here that the rates for $B_s \to K^- \pi^+$ and $\bar B_s
\to K^+ \pi^-$ can provide the necessary information on $r$, and estimate the
statistical accuracy of forthcoming measurements at the Fermilab Tevatron.
\end{quote}
\medskip

\leftline{\qquad PACS codes:  12.15.Hh, 12.15.Ji, 13.25.Hw, 14.40.Nd}
\bigskip

The measurement of phases of the Cabibbo-Kobayashi-Maskawa (CKM) matrix
\cite{CKM} is a crucial test of our understanding of CP violation.  Various
aspects of the decays $B \to K \pi$, in particular, have been shown to
provide information on the weak phase $\gamma \equiv {\rm Arg}(V^*_{ub})$
\cite{BKpi,GRKpi}.  In Ref.~\cite{GRKpi} we showed that ratios of partial
decay rates for charged and neutral $B$ mesons to $K \pi$ final states
yielded $\gamma$ when supplemented with information on the
ratio $r$ of weak tree and penguin amplitudes.  It was necessary to extract
$r$ either from data on $B \to \pi \pi$ or from semileptonic $B \to \pi$
decays.

In the present Letter we show that the decays \cite{GP}
$B_s \to K^- \pi^+$ 
and $\bar B_s \to K^+ \pi^-$, 
which are related by U-spin to  the processes $B^0 \to K^+\pi^-$ and
$\bar B^0\to K^-\pi^+$, respectively,
can provide the necessary information on $r$.  We describe the constraints
available by including information on these processes, and estimate the
statistical power of forthcoming measurements at the Fermilab Tevatron. 

We use the same procedure as Ref.~\cite{GRKpi}, which may be consulted for
further details.  However, for convenience,
we shall decompose our amplitudes into ``tree'' and ``penguin'' contributions
in a somewhat different manner.

We define the following charge-averaged ratios:
\beq
R \equiv \frac{\Gamma(B^0 \to K^+ \pi^-) + \Gamma(\bar B^0 \to K^- \pi^+)}
{\Gamma(B^+ \to K^0 \pi^+) + \Gamma(B^- \to \bar K^0 \pi^-)}~~~,
\eeq
\beq
R_s \equiv \frac{\Gamma(B_s \to K^- \pi^+) + \Gamma(\bar B_s \to K^+ \pi^-)}
{\Gamma(B^+ \to K^0 \pi^+) + \Gamma(B^- \to \bar K^0 \pi^-)}~~~,
\eeq
and CP-violating rate (pseudo-)asymmetries:
\beq
A_0 \equiv \frac{\Gamma(B^0 \to K^+ \pi^-) -\Gamma(\bar B^0 \to K^- \pi^+)}
{\Gamma(B^+ \to K^0 \pi^+) + \Gamma(B^- \to \bar K^0 \pi^-)}~~~,
\eeq
\beq
A_s \equiv \frac{\Gamma(B_s \to K^- \pi^+) - \Gamma(\bar B_s \to K^+ \pi^-)}
{\Gamma(B^+ \to K^0 \pi^+) + \Gamma(B^- \to \bar K^0 \pi^-)}~~~.
\eeq

The amplitudes for $B\to K\pi$ were expressed in Ref.~\cite{GRKpi} in terms of
``tree'' and ``penguin'' 
contributions involving CKM factors $V^*_{ub}V_{us}$ and $V^*_{tb}V_{ts}$,
respectively. Using the unitarity of the CKM matrix, it is more convenient 
in our case to decompose the amplitudes into terms containing $V^*_{ub}V_{us}$ 
and $V^*_{cb}V_{cs}$. This decomposition is in accordance with the structure 
of the $\Delta  B=1,~\Delta C=0,~\Delta S=-1$ effective Hamiltonian \cite{BBL}
\beq
{\cal H}_{\rm eff} = \frac{G_F}{\s}\left[V^*_{ub}V_{us}\left(\sum^2_1 c_i 
Q^{us}_i +\sum^{10}_3 c_i Q^s_i\right ) + V^*_{cb}V_{cs}\left(\sum^2_1 c_i 
Q^{cs}_i +\sum^{10}_3 c_i Q^s_i\right )\right]~,
\eeq
where the flavor structure of the various short-distance operators is
$Q^{qs}_{1,2}\sim\bar b q\bar q s,~Q^s_{3,..,6}\sim \bar b s \sum \bar q' q',~
Q^s_{7,..,10}\sim \bar b s\sum e_{q'}\bar q' q'$.

The amplitude of $B_s\to K^-\pi^+$ is obtained from the one for 
$B^0\to K^+\pi^-$ by a U-spin transformation, $s\leftrightarrow d$, acting 
both on the effective Hamiltonian and on the initial and final hadronic 
states. The ratios of the corresponding two CKM factors occuring in these 
amplitudes are $V^*_{ub}V_{ud}/V^*_{ub}V_{us}=1/\tl$ and $V^*_{cb}V_{cd}/
V^*_{cb}V_{cs}=-\tl$, where $\tl =|V_{us}/V_{ud}|=
|V_{cd}/V_{cs}|=\tan\theta_c\simeq 0.226$.

In a convention where the coefficient of the strangeness-changing
penguin amplitude in $B^+ \to K^0 \pi^+$ is taken to be real and positive,
we then have
\beq \label{eqn:Bplus}
A(B^+ \to K^0 \pi^+) = A(B^- \to \bar K^0 \pi^-) =P + 
{\cal O}(\frac{1}{2}\tl^2)~~~,
\eeq
$$
A(B^0 \to K^+ \pi^-) = -[P + T e^{i(\delta + \gamma)}]~~~,
$$
\beq \label{eqn:Bzero}
A(\bar B^0 \to K^- \pi^+) = - [P + T e^{i(\delta - \gamma)}]~~~,
\eeq
$$
A(B_s \to K^- \pi^+) = \tl P - (1/\tl)T e^{i(\delta + \gamma)}~~~,
$$
\beq
A(\bar B_s \to K^+ \pi^-) = \tl P - (1/\tl)T e^{i(\delta -
\gamma)}~~~,
\eeq
Here $\delta$ is the strong phase difference
between the tree and penguin amplitudes.  The ${\cal O}(\frac{1}{2}\tl^2)$ 
term in
Eq.~(\ref{eqn:Bplus}) is the relative contribution of the $V^*_{ub}V_{us}$ 
term in comparison with the dominant 
$V^*_{cb}V_{cs}$ term.  The effects of this term could be amplified if
rescattering is important.  Various estimates \cite{resc}
consider this possibility to be unlikely, but it can be checked by
measuring the CP-violating rate difference between $B^+ \to K^0 \pi^+$
and $B^- \to \bar K^0 \pi^-$ or
by improving bounds on the charge-averaged rate of the U-spin related decay
$B^{\pm} \to K^0 (\bar K^0) K^{\pm}$.
Also, when using isospin symmetry in Eqs.~(\ref{eqn:Bplus}) and 
(\ref{eqn:Bzero}) to assume 
equal penguin amplitudes in $B^+$ and $B^0$ decays, we have neglected 
color-suppressed electroweak penguin contributions \cite{GPY}. Their 
effects on determining $\gamma$ were studied in \cite{GRKpi}, and 
ways for controlling these small terms were discussed in \cite{Fl}.  

In the above equations we have taken $P$ and $T$ to be real but of
indeterminate sign.  Calculations based on the factorization hypothesis
or free-quark estimates \cite{GRKpi} suggest $T > 0$, $P < 0$, 
$\delta\approx 0$.  We define
the ratio $r \equiv T/P$. One then expects $r<0, \delta\approx 0$ in the 
factorization limit.
Note that this definition differs from the one in \cite{GRKpi},
since the present definition of $T$ contains a contribution from the
$V^*_{ub}V_{uq}$ term ($q=d,s$) in the penguin amplitude of Ref.~\cite{GRKpi}.

The charge-averaged ratios and charge asymmetries are now given by
\beq \label{eqn:R}
R = 1 + r^2 + 2 r \cos \delta \cos \gamma~~~,
\eeq
\beq \label{eqn:Rs}
R_s = \tl^2 + (r/\tl)^2 - 2 r \cos \delta \cos \gamma~~~,
\eeq
\beq \label{eqn:asym}
A_0 = - A_s = -2r \sin \gamma \sin \delta~~~.
\eeq
The expressions for $R$ and $A_0$ are those given in Ref.~\cite{GRKpi},
with a sign change in the term proportional to $r$ corresponding to our
different convention for labeling amplitudes.  The expressions for $R_s$
and $A_s$ provide new information.

Notice that since both $R_s$ and $A_s$ involve the ratios of strange and
nonstrange $B$ partial widths, their measurement in a hadronic experiment will
demand a better estimate of the relative production fraction of strange and
nonstrange $B$ mesons.  (The CDF detector at the Tevatron has measured this
ratio to be $f_s/(f_u+f_d) = 0.213 \pm 0.068$ \cite{CDF}.)
One way to accomplish this will be to compare
same-sign and opposite-sign lepton pair production \cite{SSPC}.

The equation for $R_s$ provides an opportunity to learn the magnitude of $r$, 
related to a quantity estimated previously \cite{GRKpi} with the help of 
$B \to \pi \pi$
or $B \to \pi \ell \nu_\ell$ decays.  Adding Eqs.~(\ref{eqn:R}) and
(\ref{eqn:Rs}), we find
\beq \label{eqn:r}
R + R_s = (1+ \tl^2)(1 + \frac{r^2}{\tl^2})~~,~~~
{\rm or}~~
|r| = \tl \left[ \frac{R+R_s}{1+\tl^2} - 1 \right]^{1/2}~~~.
\eeq

Once $|r|$ is known we can establish a useful bound on $\gamma$
independent of any possible CP-violating charge-asymmetry.  We write
\beq
(R - 1 - r^2)^2 = 4 r^2 \cos^2 \delta \cos^2 \gamma \le 4 r^2 \cos^2 \gamma
~~~,
\eeq
implying 
\beq
|\cos\gamma| \ge \frac{|R-1-r^2|}{2|r|}~,
\eeq
or
\beq
|\sin \gamma| \le \frac{[-\lambda(1,R,r^2)]^{1/2}}{2|r|}~~,~~~
\lambda(a,b,c) \equiv a^2+b^2+c^2 -2ab - 2ac -2bc~~~.
\eeq

The limiting cases of an equality, in which $\gamma$ is determined up to
a twofold ambiguity, $\gamma \to \pi -\gamma$, correspond to $\delta=0$ or 
$\delta=\pi$. Assuming that the strong phase difference between $T$ and $P$
is small, as indicated by perturbative QCD  arguments 
\cite{BSS,BBNS}, measurements of $R$ and $R_s$ already determine $\gamma$.
Note that we expect $r<0$ for a vanishing final-state phase $\delta$,
in which case one has
\beq
\cos\gamma = \frac{1 + r^2 - R}{2|r|}~.
\eeq
In case that the strong phase is large, one expects to measure nonzero 
asymmetries $A_0=-A_s$, which together with $R$ and $R_s$ would determine 
$\gamma$.
As quoted previously \cite{GRKpi}, by eliminating $\delta$ from the
equations for $R$ and $A_0$, one finds
\beq
R = 1 + r^2 \pm \sqrt{4r^2 \cos^2 \gamma - A_0^2 \cot^2 \gamma}~~~,
\eeq
which can then be solved for $\sin^2 \gamma$ up to a two-fold ambiguity
when $A_0 \ne 0$.

We have treated the systematic effects associated with theoretical
uncertainties in our method in Ref.~\cite{GRKpi}.  There it was shown
that an error on $\gamma$ of about $10^\circ$ could be achieved if one
obtained an error of about 10\% on $r$ 
(defined in a somewhat different manner) for the value favored there,
$r = 0.16$.  When obtaining $r$ using the measurement of $R$ and $R_s$,
the statistically limiting measurement is likely to be that of $R_s$,
since the $B_s$ is expected to be produced less copiously than the $B^0$,
and the tree amplitude $T/\tl$ dominating the decay
$B_s \to K^- \pi^+$ is expected to be smaller than the amplitude $P$ dominating
the decay $B^0 \to K^+ \pi^-$.  In one estimate \cite{Wurt,Jesik}, the
ratio of produced $B_s \to K^- \pi^+$ : $B^0 \to K^+ \pi^-$ events is
expected to be 1:8.   

As one example, we take $R=1$ (the present experimental value is $1.01 \pm
0.30$ \cite{CLEO}), and find that $r = 0.16 \pm 0.016$ corresponds to
$R_s = 0.58^{+0.11}_{-0.10}$.  Thus, one must measure $R_s$ to about
$\pm 18 \%$ in order to achieve a 10\% measurement of $r$.  This would require
about 30 events of $B_s \to K^- \pi^+$ or $\bar B_s \to K^+ \pi^-$,
whereas a sample of about 2000 to 4000 untagged events,
two orders of magnitude larger,
is envisioned for Run II at the Tevatron \cite{Wurt,Jesik}.

It is sufficient to use untagged events since the rapid $B_s$--$\bar B_s$
mixing ensures that the effective production rates for $B_s \to K^- \pi^+$ and
$\bar B_s \to K^+ \pi^-$ will be very nearly identical
at each value of rapidity.  While this is not so for $B^0 \to K^+\pi^-$ and
$\bar B^0 \to K^- \pi^+$, it is true in proton-antiproton collisions if one
averages over rapidity in a symmetric detector.  Otherwise,
one must know the individual production rates for
$B^0$ and $\bar B^0$ {\it before mixing}. These can be measured, for example
using self-tagging processes such as $B^0 \to J/\psi K^{*0}$ where the
CP-violating asymmetry is expected to be small, or with the help of an
assumption of isospin invariance and a corresponding measurement of $B^+$
and $B^-$ production using self-tagging modes with small CP asymmetry
such as $B^\pm \to J/\psi K^\pm$ \cite{CLEOB}.
  
In order to estimate the systematic error associated with our flavor
SU(3) approximation, we adopt the often used assumption of factorized 
tree and penguin amplitudes.  In this approximation, SU(3) breaking 
introduces in the ratio $A(B_s\to K^-\pi^+)/A(B^0\to K^+\pi^-)$ a relative 
factor $f_1=F_{B_s K}(m^2_\pi)f_{\pi}/F_{B\pi}(m^2_K)f_K$, involving
the product of corresponding ratios of form factors and decay constants 
describing $B_s\to K\pi$ and $B\to K\pi$.
Although $B$ decay form factors to $K$ and $\pi$ have not yet 
been measured, their ratio is expected to be somewhat larger than one; quark 
model \cite{BSW} and QCD sum rule \cite{Ball} 
calculations find $F_{B_s K}(m^2_\pi)/F_{B\pi}(m^2_K)\simeq 1.16$. With  
$f_K/f_{\pi}=1.22$, SU(3) breaking is expected to introduce a systematic 
uncertainty of only a few percent, $f_1\simeq 0.95$.

An actual check of the SU(3) (or U-spin) approximation relies on the asymmetry 
prediction $A_s = - A_0$, or, in fact, on the prediction of equal CP-violating
rate differences in $B^0\to K\pi$ and  $B_s\to K\pi$. 
Including SU(3) breaking, one expects
\beq
\Gamma(B_s\to K^-\pi^+) - \Gamma(\bar B_s\to K^+\pi^-)
= -f^2_1[\Gamma(B^0\to K^+\pi^-) - \Gamma(\bar B^0\to K^-\pi^+)]~.
\eeq 
With 2000 events of $B_s \to K^- \pi^+$ or $\bar B_s \to K^+ \pi^-$,
one should be able to check this prediction to a fractional accuracy of
$(2000)^{-1/2} \simeq 2.2\%$ which seems adequate
for a precise determination of $f_1$. This factor can then be included in
the analysis leading to a determination of $\gamma$.
Note, however, that a very small 
strong phase difference $\delta$ may not permit useful asymmetry measurements. 
(The strong phases in $B^0\to K\pi$ and $B_s\to K\pi$ differ in broken SU(3), 
in spite of the fact that the final states in both decays are $K^\pm\pi^\mp$. 
This is similar to the case of $K \pi$ states produced in Cabibbo-favored and 
double-Cabibbo-suppressed $D$ decays \cite{FNP}.)

The processes studied here, $B^0\to K^+\pi^-$ and $B_s\to K^-\pi^+$, 
are related by a spectator quark transformation, $d\leftrightarrow s$, to 
another pair of U-spin related processes $B_s\to K^+ K^-$ and
$B^0\to \pi^+\pi^-$.
The amplitudes of these two pairs of processes are equal, respectively, 
in the limit of flavor SU(3) symmetry and neglecting smaller ``exchange" and
``penguin annihilation" amplitudes \cite{GHLR}. These contributions 
are expected to be very small unless rescattering is important
\cite{resc}. This can be tested by measuring the rates for $B^0\to K^+K^-$
or $B_s\to \pi^+\pi^-$ which are due solely to exchange and penguin annihilation
amplitudes. 
Including an SU(3) breaking factor 
$f_2=F_{B_s K}(m^2_K)/F_{B\pi}(m^2_K)\approx 
F_{B_s K}(m^2_{\pi})/F_{B\pi}(m^2_{\pi}) \simeq 1.16$, as expected in the 
factorization approximation, one has
$$
A(B_s\to K^+ K^-) = f_2 A(B^0\to K^+\pi^-)~,
$$
\beq\label{su3}
A(B^0\to \pi^+\pi^-) = f^{-1}_2 A(B_s\to K^-\pi^+)~.
\eeq
The rates of these four processes can be used to 
check the factorization assumption and to 
determine the SU(3) breaking factor $f_2$. A study of $\gamma$ through 
time-integrated rates of $B_s\to K^+ K^-$, 
$B^0\to \pi^+\pi^-$ and their charge-conjugates 
is very similar and complementary to the one 
using $B^0/\bar B^0\to K^{\pm}\pi^{\mp}$ and $B_s/\bar B_s\to K^{\mp}\pi^{\pm}$.

The present proposal for determining $\gamma$ is also complementary to the 
method comparing time-dependent asymmetry measurements
in $B^0 \to \pi^+ \pi^-$ and in $B_s \to K^+ K^-$ \cite{Wurt,RF}.  
In that method, one measures
both an oscillation amplitude and an oscillation phase for both channels,
i.e., four quantities, and thereby extracts four unknowns:  the ratio of
tree and penguin amplitudes (equivalent to our $r$), a strong final-state
phase $\delta$ between tree and penguin assumed to be the same for $\pi^+
\pi^-$ and $K^+ K^-$, and two weak phases $\beta$ and $\gamma$.  One must
tag the flavor of the neutral mesons at time of production.
 
We comment briefly on the relation between the two methods. The
present method has the following advantages:

(1) It requires neither time-dependent measurements nor flavor
tagging for $B^0$ and $B_s$.

(2) The kinematic peaks associated with $B^0 \to K \pi$ and $B_s \to K \pi$
are well-separated from one another and reasonably well separated from those
corresponding to $B^0 \to \pi^+ \pi^-$ and $B_s \to K^+ K^-$ (which lie
between them and are nearly on top of each other), even without particle 
identification \cite{Wurt,Jesik}. To see this,
we note that in the limit of relativistic pions or kaons, the mass $m_{21}$
of a $\pi \pi$, $K \pi$, or $K \bar K$ system is given by
\beq
m_{12} = \left[ m_1^2 \left( 1 + \frac{p_2}{p_1} \right)
       +        m_2^2 \left( 1 + \frac{p_1}{p_2} \right)
       + 2 p_1 p_2(1 - \cos \theta_{12}) \right]^{1/2}~~~,
\eeq
where $p_1$ and $p_2$ are laboratory momenta of the two final-state tracks
and $\theta_{12}$ is the laboratory angle between them.  If we were to call
both tracks pions and neglect $m_1^2$ and $m_2^2$, we would thus estimate
\beq
(m^{\rm est}_{12})^2 = 2 p_1 p_2(1 - \cos \theta_{12})
= m_{12}^2 -  m_1^2 \left( 1 + \frac{p_2}{p_1} \right) 
       +      m_2^2 \left( 1 + \frac{p_1}{p_2} \right) ~~~.
\eeq
For the case of symmetric momenta $p_1 = p_2$ corresponding to the
Jacobian peak of the distribution, we would thus estimate
\beq
m_{12}^{\rm est} \simeq (m_{12} - 2 m_1^2 - 2 m_2^2)^{1/2}~~~,
\eeq
or about 45 MeV below the $B$ for $B \to K \pi$, $\simeq
M_B$ for both $B \to \pi \pi$
and $B_s \to K \bar K$, and about 45 MeV above the $B$ for $B_s \to K \pi$.

The CDF Detector at Run II of the Tevatron will have a time-of-flight
system for particle identification, whose capabilities of distinguishing
kaons from pions in $B$ decay will be minimal.  It will also make use
of $dE/dx$ discrimination.  Later-generation detectors,
such as LHC-b at CERN and the proposed BTeV detector at Fermilab, will have
better capabilities for particle identification, allowing for cleaner
separation of $\pi \pi$, $K \pi$, and $K \bar K$ final states of $B$ and
$B_s$ decay.

The method based on $B^0 \to \pi^+ \pi^-$ and $B_s \to K^+
K^-$ has the following points in its favor:

(1) It involves no small corrections from rescattering and color-suppressed
electroweak penguin contributions. Also, certain ratios of form factors and 
decay constants cancel one another
in the factorization limit, since one is only discussing asymmetries.

(2) Although the peaks for $B \to \pi \pi$ and $B_s \to K \bar K$ lie nearly
on top of one another in the absence of particle identification, the
time-dependent CP-violating rate differences in the two cases can be separated 
from one another (as long as they both are nonzero) by a frequency analysis. 
For the two asymmetries one would still need particle identification in order
to distinguish between the denominators defining $B^0$ and $B_s$ asymmetries.
  
(3) The analysis also allows for a lifetime difference between CP-even and
CP-odd $B_s$ mass eigenstates \cite{RF}, which may allow
resolution of some discrete ambiguities present in our scheme \cite{GRKpi}. 

To conclude, we have found that simultaneous measurement of rates
for $B^0 \to K^+ \pi^-$, $B_s \to K^- \pi^+$, and their charge-conjugates
can provide useful information on the weak phase $\gamma$.  This measurement
appears feasible at the Fermilab Tevatron and is an appealing possibility
for other facilities envisioning $B_s$ production, such as the HERA-B
detector at DESY, the BTeV detector at Fermilab, and the LHC-b detector
at CERN.  Effects of SU(3) breaking and rescattering appear to be controllable
and to some extent testable, while electroweak penguin effects are expected
to be small enough that a determination of $\gamma$ to better than $10^\circ$
appears feasible.  Our method, which can also be applied to time-integrated
rates of  $B^0 \to \pi^+ \pi^-$ and $B_s \to K^+ K^-$, appears complementary to
that \cite{Wurt,RF} based on time-dependent asymmetries in the latter
processes, for which comparable accuracy in $\gamma$ is expected \cite{Wurt}.
Whereas the latter determination of $\gamma$ could be affected by new physics
contributions to $B_s-\bar B_s$ mixing \cite{GL}, our method is free of such 
modifications.

We thank S. Stone and F. W\"urthwein for useful discussions. 
M. G. would like to acknowledge the very kind hospitality of the Enrico Fermi 
Institute at the University of Chicago.
This work was supported in part by the United States -- Israel Binational
Science Foundation under 
Research Grant Agreement 98-00237 and by the United States Department of 
Energy under Contract No.~DE-FG02-90-ER40560. 
\bigskip

\def \ajp#1#2#3{Am.~J.~Phys.~{\bf#1} (#3) #2}
\def \apny#1#2#3{Ann.~Phys.~(N.Y.) {\bf#1} (#3) #2}
\def \app#1#2#3{Acta Phys.~Polonica {\bf#1} (#3) #2 }
\def \arnps#1#2#3{Ann.~Rev.~Nucl.~Part.~Sci.~{\bf#1} (#3) #2}
\def \cmp#1#2#3{Commun.~Math.~Phys.~{\bf#1} (#3) #2}
\def \cmts#1#2#3{Comments on Nucl.~Part.~Phys.~{\bf#1} (#3) #2}
\def \cn{Collaboration}
\def \corn93{{\it Lepton and Photon Interactions:  XVI International Symposium,
Ithaca, NY August 1993}, AIP Conference Proceedings No.~302, ed.~by P. Drell
and D. Rubin (AIP, New York, 1994)}
\def \cp89{{\it CP Violation,} edited by C. Jarlskog (World Scientific,
Singapore, 1989)}
\def \dpff{{\it The Fermilab Meeting -- DPF 92} (7th Meeting of the American
Physical Society Division of Particles and Fields), 10--14 November 1992,
ed. by C. H. Albright \ite~(World Scientific, Singapore, 1993)}
\def \dpf94{DPF 94 Meeting, Albuquerque, NM, Aug.~2--6, 1994}
\def \efi{Enrico Fermi Institute Report No. EFI}
\def \el#1#2#3{Europhys.~Lett.~{\bf#1} (#3) #2}
\def \epjc#1#2#3{Eur.~Phys.~J.~C~{\bf#1} (#3) #2}
\def \f79{{\it Proceedings of the 1979 International Symposium on Lepton and
Photon Interactions at High Energies,} Fermilab, August 23-29, 1979, ed.~by
T. B. W. Kirk and H. D. I. Abarbanel (Fermi National Accelerator Laboratory,
Batavia, IL, 1979}
\def \hb87{{\it Proceeding of the 1987 International Symposium on Lepton and
Photon Interactions at High Energies,} Hamburg, 1987, ed.~by W. Bartel
and R. R\"uckl (Nucl. Phys. B, Proc. Suppl., vol. 3) (North-Holland,
Amsterdam, 1988)}
\def \ib{{\it ibid.}~}
\def \ibj#1#2#3{~{\bf#1} (#3) #2}
\def \ichep72{{\it Proceedings of the XVI International Conference on High
Energy Physics}, Chicago and Batavia, Illinois, Sept. 6--13, 1972,
edited by J. D. Jackson, A. Roberts, and R. Donaldson (Fermilab, Batavia,
IL, 1972)}
\def \ijmpa#1#2#3{Int.~J.~Mod.~Phys.~A {\bf#1} (#3) #2}
\def \ite{{\it et al.}}
\def \jhep#1#2#3{JHEP~{\bf#1} (#3) #2}
\def \jmp#1#2#3{J.~Math.~Phys.~{\bf#1} (#3) #2}
\def \jpg#1#2#3{J.~Phys.~G {\bf#1} (#3) #2}
\def \lkl87{{\it Selected Topics in Electroweak Interactions} (Proceedings of
the Second Lake Louise Institute on New Frontiers in Particle Physics, 15--21
February, 1987), edited by J. M. Cameron \ite~(World Scientific, Singapore,
1987)}
\def \ky85{{\it Proceedings of the International Symposium on Lepton and
Photon Interactions at High Energy,} Kyoto, Aug.~19-24, 1985, edited by M.
Konuma and K. Takahashi (Kyoto Univ., Kyoto, 1985)}
\def \mpla#1#2#3{Mod.~Phys.~Lett.~A {\bf#1} (#3) #2}
\def \nc#1#2#3{Nuovo Cim.~{\bf#1} (#3) #2}
\def \np#1#2#3{Nucl.~Phys.~{\bf#1} (#3) #2}
\def \pisma#1#2#3#4{Pis'ma Zh.~Eksp.~Teor.~Fiz.~{\bf#1} (#3) #2[JETP Lett.
{\bf#1} (#3) #4]}
\def \pl#1#2#3{Phys.~Lett.~{\bf#1} (#3) #2}
\def \plb#1#2#3{Phys.~Lett.~B {\bf#1} (#3) #2}
\def \pr#1#2#3{Phys.~Rev.~{\bf#1} (#3) #2}
\def \pra#1#2#3{Phys.~Rev.~A {\bf#1} (#3) #2}
\def \prd#1#2#3{Phys.~Rev.~D {\bf#1} (#3) #2}
\def \prl#1#2#3{Phys.~Rev.~Lett.~{\bf#1} (#3) #2}
\def \prp#1#2#3{Phys.~Rep.~{\bf#1} (#3) #2}
\def \ptp#1#2#3{Prog.~Theor.~Phys.~{\bf#1} (#3) #2}
\def \rmp#1#2#3{Rev.~Mod.~Phys.~{\bf#1} (#3) #2}
\def \rp#1{~~~~~\ldots\ldots{\rm rp~}{#1}~~~~~}
\def \si90{25th International Conference on High Energy Physics, Singapore,
Aug. 2-8, 1990}
\def \slc87{{\it Proceedings of the Salt Lake City Meeting} (Division of
Particles and Fields, American Physical Society, Salt Lake City, Utah, 1987),
ed.~by C. DeTar and J. S. Ball (World Scientific, Singapore, 1987)}
\def \slac89{{\it Proceedings of the XIVth International Symposium on
Lepton and Photon Interactions,} Stanford, California, 1989, edited by M.
Riordan (World Scientific, Singapore, 1990)}
\def \smass82{{\it Proceedings of the 1982 DPF Summer Study on Elementary
Particle Physics and Future Facilities}, Snowmass, Colorado, edited by R.
Donaldson, R. Gustafson, and F. Paige (World Scientific, Singapore, 1982)}
\def \smass90{{\it Research Directions for the Decade} (Proceedings of the
1990 Summer Study on High Energy Physics, June 25 -- July 13, Snowmass,
Colorado), edited by E. L. Berger (World Scientific, Singapore, 1992)}
\def \stone{{\it B Decays}, edited by S. Stone (World Scientific, Singapore,
1994)}
\def \tasi90{{\it Testing the Standard Model} (Proceedings of the 1990
Theoretical Advanced Study Institute in Elementary Particle Physics, Boulder,
Colorado, 3--27 June, 1990), edited by M. Cveti\v{c} and P. Langacker
(World Scientific, Singapore, 1991)}
\def \yaf#1#2#3#4{Yad.~Fiz.~{\bf#1} (#3) #2 [Sov.~J.~Nucl.~Phys.~{\bf #1} (#3)
#4]}
\def \zhetf#1#2#3#4#5#6{Zh.~Eksp.~Teor.~Fiz.~{\bf #1} (#3) #2 [Sov.~Phys. -
JETP {\bf #4} (#6) #5]}
\def \zpc#1#2#3{Zeit.~Phys.~C {\bf#1} (#3) #2}


\begin{thebibliography}{99}

\bibitem{CKM} N. Cabibbo, \prl{10}{531}{1963}; M. Kobayashi and T. Maskawa,
\ptp{49}{652}{1973}.

\bibitem{BKpi} M. Gronau, J. Rosner and D. London, \prl{73}{21}{1994};
R. Fleischer, \plb{365}{399}{1996}; \prd{58}{093001}{1998};
M. Gronau and J. L. Rosner, \prl{76}{1200}{1996};
A. S. Dighe, M. Gronau, and J. L. Rosner, \prd{54}{3309}{1996};
A. S. Dighe and J. L. Rosner, \ibj{54}{4677}{1996};
R. Fleischer and T. Mannel, \prd{57}{2752}{1998};
F. W\"urthwein and P. Gaidarev, hep-ph/9712531, December, 1997 (unpublished);
M. Neubert and J. L. Rosner, \plb{441}{403}{1998}; \prl{81}{5076}{1998};
A. J. Buras, R. Fleischer, and T. Mannel, \np{B533}{3}{1998};
R. Fleischer and A. J. Buras, \epjc{11}{93}{1999};
M. Neubert, \jhep{9902}{014}{1999};
M. Gronau and D. Pirjol, \prd{61}{013005}{2000}.

\bibitem{GRKpi} M. Gronau and J. L. Rosner, \prd{57}{6843}{1998}.

\bibitem{GP} M. Gronau and D. Pirjol, \plb{449}{321}{1999}.

\bibitem{BBL} G. Buchalla, A. J. Buras, and M. E. Lautenbacher, 
\rmp{68}{1125}{1996}.

\bibitem{resc} B. Blok, M. Gronau, and J. L. Rosner, \prl{78}{3999}{1997};
\ibj{79}{1167}{1997};
A. Falk, A. L. Kagan, Y. Nir, and A. A. Petrov, \prd{57}{4290}{1998};
M. Gronau and J. L. Rosner, \prd{57}{6843}{1998}; \ibj{58}{113005}{1998};
R. Fleischer, \plb{435}{221}{1998}; \epjc{6}{451}{1999}.

\bibitem{GPY} M. Gronau, D. Pirjol and T. -M. Yan, \prd{60}{034021}{1999}.

\bibitem{Fl} R. Fleischer, \epjc{6}{451}{1999}.

\bibitem{CDF} CDF \cn, T. Affolder \ite, \prl{84}{1663}{2000}.

\bibitem{SSPC} S. Stone, private communication.

\bibitem{BSS} M. Bander, D. Silverman, and A. Soni, \prl{43}{242}{1979}.

\bibitem{BBNS} M. Beneke, G. Buchalla, M. Neubert and C. Sachrajda, 
\prl{83}{1914}{1999}.

\bibitem{Wurt} F. W\"urthwein, talk for Working Group 1 presented at
Workshop on B Physics at the Tevatron -- Run II and Beyond, Fermilab, February 
2000 (unpublished).

\bibitem{Jesik} R. Jesik, summary talk for Working Group 1 presented at
Workshop on B Physics at the Tevatron -- Run II and Beyond, Fermilab, February 
2000 (unpublished).

\bibitem{CLEO} CLEO Collaboration, D. Cronin-Hennesy \ite, Cornell University
report CLNS 99/1650, hep-ex/0001010, unpublished.

\bibitem{CLEOB} CLEO Collaboration, G. Bonvicini \ite, Cornell University
report CLNS 00/1661, hep-ex/0003004, unpublished.

\bibitem{BSW} M. Bauer, B. Stech and M. Wirbel, \zpc{29}{637}{1985}.

\bibitem{Ball} P. Ball, \jhep{9809}{005}{1998}.

\bibitem{FNP} A. F. Falk, Y. Nir, and A. A. Petrov, \jhep{9912}{019}{1999}.

\bibitem{GHLR} M. Gronau, O. F. Hern\'andez, D. London and J. L. Rosner,
\prd{50}{4529}{1994}.

\bibitem{RF} R. Fleischer, \plb{459}{306}{1999};
DESY preprint DESY 00-014, hep-ph/0001253. See also I. Dunietz, Proceedings of 
the Workshop on $B$ Physics at Hadron Accelerators, Snowmass, CO, 1993, p. 83;
D. Pirjol, \prd{60}{054020}{1999}.

\bibitem{GL} Y. Nir and D. Silverman, \np{B345}{301}{1990}; C. O. Dib, D. 
London and Y. Nir, \ijmpa{6}{1253}{1991}; M. Gronau and D. London, 
\prd{55}{2845}{1997}.
\end{thebibliography}
\end{document}